\documentclass[%
 aip,
 amsmath,amssymb
 reprint,%
]{revtex4-1}
\usepackage{graphicx}
\usepackage{dcolumn}
\usepackage{bm}

\usepackage[dvipsnames]{xcolor}
\usepackage{soul}
\usepackage[utf8]{inputenc}
\usepackage[T1]{fontenc}
\usepackage{mathptmx}
\usepackage{etoolbox}


\begin{document}

\preprint{AIP/123-QED}

\title{Thermal-bioconvection in a non-scattering suspension of phototactic microorganisms}
\author{Sandeep Kumar}
\thanks{Corresponding author}
\email{sandeepkumar1.00123@gmail.com}
\affiliation{
Department of Mathematics, PDPM  Indian Institute of Information Technology Design and Manufacturing, Jabalpur 482005, India
}%

\author{Shaowei Wang}
\email{shaoweiwang@sdu.edu.cn}
\affiliation{
Department of Engineering Mechanics, School of Civil Engineering, Shandong University, Jinan 250061, PR China
}


\date{\today}

\begin{abstract}

This article investigates the linear stability of thermal-bioconvection within a suspension containing phototactic microorganisms heated from below. In suspension, the upper surface is taken as stress-free, while the lower surface is taken as rigid. The resulting eigenvalue problem, including the bioconvection Rayleigh and thermal Rayleigh numbers, is resolved numerically. Changes in the critical total intensity and Lewis number do not impact the critical threshold of the thermal Rayleigh number; however, they notably influence the critical bioconvection Rayleigh number. The critical total intensity and Lewis number destabilize the suspension. It is observed that heating from below enhances the instability of the layer. At higher temperatures, Rayleigh-B$\acute{e}$nard convection dominates bioconvection, resulting in a single convection cell.

\end{abstract}

\maketitle

\section{Introduction}
The process known as bioconvection is one in which swimming microorganisms in suspensions produce patterns that are not predetermined \cite{ref1}. Wager \cite{ref2} was the first author to conduct the initial extensive experiments pertaining to the research of bioconvection, while Platt \cite{ref3} was the one who came up with the term ``bioconvection''. The microorganisms are slightly denser than the fluid and aggregate due to tactic behavior; however, microorganisms have a tendency to move upward, which results in the development of patterns. If microorganisms cease swimming, patterns are no longer visible. The formation of a pattern does not always require swimming in an upward direction or reaching a greater density \cite{ref1}. Pattern types may be influenced by depth, the concentration of microorganisms, and motion. Pattern creation may be seen in the flagellated green algae \textit{Volvox}, \textit{Euglena}, \textit{Dunaliella}, and \textit{Chlamydomonas}, among others \cite{ref6,ref5,ref4,ref2}. The reactions of microorganisms to external stimuli are referred to collectively as taxes. Some examples of taxes are chemotaxis, gravitaxis, gyrotaxis, and phototaxis. Chemotaxis is the response of an organism to changes in chemical gradients, whereas gravitaxis is the response of an organism to changes in the force of gravity. The torque due to gravity and shear is balanced to generate gyrotaxis. The concept of phototaxis, which is brought up in this piece, refers to motion in either the direction of or away from light. Experiments provide information on the roles that light intensity and gradient play in determining bioconvection patterns \cite{ref5,ref2}. The pattern's contour, size, and scale change depending on the amount of light \cite{ref7}. The total light intensity $\mathcal{G}$ in comparison to a critical value $\mathcal{G}_c$ and the amount of light absorbed by microorganisms are both factors that cause pattern variations \cite{ref8}. 

In a suspension, which is lighted from above and has a finite depth, the fundamental equilibrium state (the steady state) is defined by an equilibrium between the phototaxis that is caused by diffusion and the absorption of light. As a direct result of this, a sublayer is horizontal and densely inhabited by microorganisms. The region above this sublayer maintains its gravitational stability, in contrast to the region that lies underneath it, which undergoes gravitational instability. The critical total intensity is denoted by the symbol $\mathcal{G}_c$, which affects the location of the sublayer. If, across the whole of the suspension, the total intensity $\mathcal{G}$ is found to be less than $\mathcal{G}_c$, the sublayer will develop at the top of the suspension. On the other hand, if $\mathcal{G}$ is found to be larger, the sublayer will form at the bottom boundary. When the critical total intensity $\mathcal{G}_c$ is equal to the intensity $\mathcal{G}$ of the suspension as a whole, the location of the sublayer is in the middle of the upper and lower bounds. Therefore, if an unstable fluid layer were to arise, its movement would extend into the stable layer through a process known as penetrative convection \cite{ref10}.

The majority of early studies focused on suspensions in fluids at a constant temperature. In contrast, many microbes, such as the thermophilic ones found in hot springs, flourish in environments with wide temperature swings. Several authors have conducted extensive research in the domain of bio-thermal convection. Kuznetsov \cite{kuznetsov2005onset} studied the instability of gyrotactic microorganisms heated from below in suspensions. A linear stability analysis technique demonstrated a relationship between bioconvection and natural convection. Kuznetsov\cite{kuznetsov2011non} investigated the biothermal convection in a suspension that included both gyrotactic microorganisms and nanoparticles. Both stationary and oscillatory kinds of convection were investigated in this study. Also, the thermal Rayleigh number dependent on nanoparticles and the bioconvection Rayleigh number were investigated. Zhao et al. \cite{zhao2018linear} used a random swimming model to investigate the linear stability of bioconvection of gyrotactic microorganisms heated from below. An investigation into a particular situation characterized by non-oscillatory instability and two rigid boundary conditions was carried out by the authors. Using the Darcy-Brinkman model, Zhao et al. \cite{zhao2019darcy} conducted an exploration into the phenomenon of bio-thermal convection within a highly porous medium, where the heat source was applied from beneath. Balla et al. \cite{balla2020bioconvection} conducted a study on the phenomenon of bioconvection in a square enclosure with porous walls, focusing on the behavior of oxytactic microorganisms under the influence of thermal radiation. Inside a vertical wavy porous cavity, Hussain et al. \cite{hussain2022thermal} investigated the effect of thermal radiation on the bioconvection flow of nano-enhanced phase change materials and oxytactic microorganisms.

In the beginning, the idea of phototactic bioconvection was originally presented by Vincent and Hill  \cite{ref9}, who explored a model in non-scattering absorbing suspension. The stability of a suspension of phototactic microorganisms was investigated using this model of phototaxis and shading. The suspension was uniformly illuminated from above, and the microorganisms were swimming in a fluid that was somewhat less dense. After that, Ghorai and Hill \cite{ref13} devised a two-dimensional phototactic model and investigated the stability of their creation. To numerically solve the governing equations, they resorted to a conservative form of the finite-difference approach. Nevertheless, the scattering effects were not taken into account in these models. Later, Ghorai et al.  \cite{ref14} updated the phototactic bioconvection model to include isotropic scattering. The new model indicated that owing to dispersion, microorganisms congregate in two horizontal layers at differing depths whenever certain parameter values are applied. In an extension of the research done by Ghorai et al. \cite{ref14}, Kumar \cite{kumar2023} showed that the system was more stable when the top surface was rigid as compared to the stress-free surface. The effects of forward anisotropic scattering on phototactic bioconvection suspension were investigated by Ghorai and
Panda \cite{ref15}. In their computational research, the impacts of forward scattering became the primary emphasis. After that, Panda et al. \cite{ref16}, Panda \cite{panda2020effects} looked at how a scattering suspension was affected by both diffuse and collimated light. They displayed a wide variety of behaviors in response to the various conditions. In addition, Panda et al. \cite{ref17} investigated a phototactic bioconvection model using oblique light and a non-scattering suspension. They were able to observe transitions between stable and overstable modes for a variety of angles of incidence and particular parameter variations. An investigation was conducted by Kumar \cite{ref18}, Panda and Rajput \cite{panda2023phototactic} to determine how oblique irradiation affects an isotropic scattering suspension. As the angle of incidence was changed, varied scattering albedo values resulted in distinct patterns for the concentration profile. None of these models of phototactic bioconvection took into account the effect that rotation might have. Recently, Kumar \cite{kumar2023effect} first time investigated the effect of rotation on the non-scattering medium and analyzed the linear instability. In this model, the author illustrated that rotation has a stabilizing effect on the suspension. Despite the fact that these observations are intriguing, there has been no theoretical research that takes into account both the phototactic and temperature gradient processes.

In the present study, we will make use of the phototaxis model that Vincent and Hill \cite{ref9} put out. The Navier-Stokes equation in conjunction with a cell conservation equation to accurately model an incompressible fluid is utilized. The phototactic suspension, according to our hypothesis, is heated from below and illuminated from above via the use of vertical collimated irradiation. A significant amount of motile algae are dependent on photosynthesis for their source of nutrition, which causes them to exhibit phototaxis behavior. It is essential to investigate a phototaxis model inside a thermal medium to convey this model realistically.

The paper is structured in a subsequent manner: In Section \ref{sec2}, the mathematical model for phototactic bioconvection is presented, along with the corresponding boundary conditions. The steady-state solution is presented in section \ref{sec3}, followed by a linear stability analysis in section \ref{sec4}. Subsequently, the numerical results and conclusion are explained in sections \ref{sec5} and \ref{sec6}.

\begin{figure*}
    \centering
    \includegraphics[width=16cm, height=10cm ]{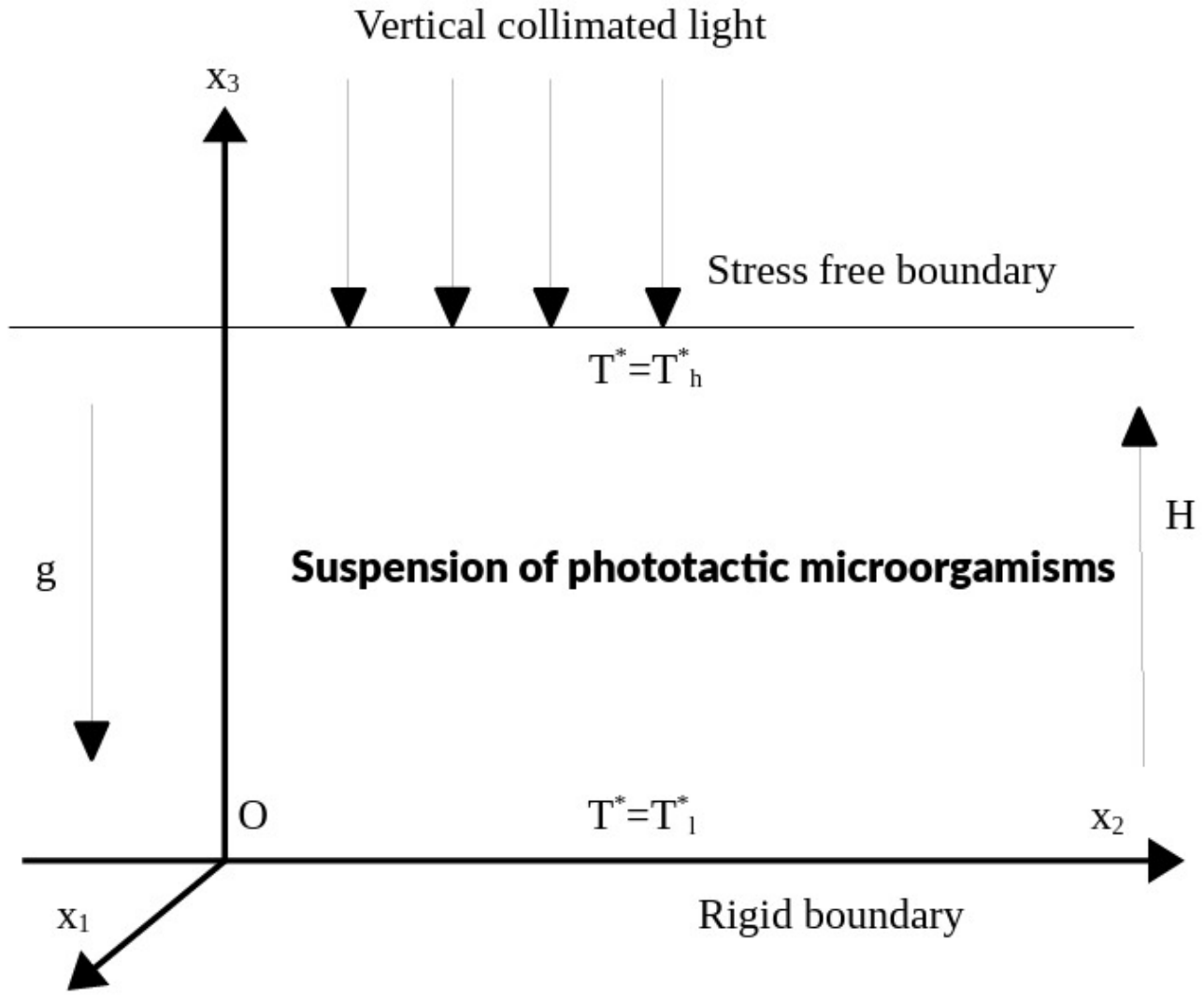}
    \caption{The geometry of the problem. }
   \label{fig:thermalfig.eps}
 \end{figure*}

\section{MATHEMATICAL MODEL}
\label{sec2}
Take into consideration a Cartesian coordinate system for three dimensions that is referred to as $O x_1 x_2 x_3$. This system extends to an infinite width, although it only has a finite depth, denoted by $H$. The lower boundary, at the point where $x_3=0$, is considered rigid, whereas the upper boundary at $x_3=H$, appears as stress-free (free from stress). At the top boundary, there is direct collimated lighting emanating from a source of light. We devote particular attention to the effect that collimated illumination contains, and we are ignoring any impacts that may be caused by light that is angled or dispersed. In addition, we are not taking into account the impact that rotation has on the suspension. We use the function $I(\boldsymbol{x},\boldsymbol{r})$ to describe the intensity at a certain point $\boldsymbol{x}=(x_1,x_2,x_3)$ in a particular direction represented as $\boldsymbol{r}$. This function describes the intensity as having components $\sin{\alpha_1}\cos{\alpha_2}$ in the $\boldsymbol{i}$ direction, $\sin{\alpha_1}\sin{\alpha_2}$ in the $\boldsymbol{j}$ direction, and $\cos{\alpha_1}$ in the $\boldsymbol{k}$ direction. In this scenario, the polar angle is denoted by $\alpha_1$, while the azimuthal angle is denoted by $\alpha_2$.

The continuity equation is given by
\begin{equation}
\label{eqn:equation8}
    div(\boldsymbol{v})=0.
\end{equation}

The equation for momentum, related to the Boussinesq, can be written as \cite{zhao2018linear}
\begin{align}
    \label{eqn:equation9}
    \varrho\left(\frac{\partial}{\partial t}+\boldsymbol{v}\cdot\nabla\right)\boldsymbol{v}&=\mu\nabla^2\boldsymbol{v}-\boldsymbol{\nabla} \mathcal{P}-n\vartheta \Delta \varrho g \boldsymbol{k}  -\varrho (1-\beta(T^*-T_u^*))g \boldsymbol{k}.
\end{align}

In this context, time is denoted by $t$, fluid velocity by $\boldsymbol{v}$, dynamic viscosity by $\mu$, fluid density by $\rho$, the density of each algal cell by $\rho+\Delta \rho$ $(\Delta \rho/\rho \ll 1)$, volume of each algal cell by $\vartheta$, gravitational acceleration by $g$, cell concentration by $n$, excess pressure above hydrostatic by $\mathcal{P}$, volumetric thermal expansion coefficient by $\beta$. The temperature of the fluid is indicated by $T^*$, while the temperature of the upper wall is represented by $T_u^*$ and the temperature of the lower wall is represented by $T_l^*$ accordingly.

The thermal energy equation is given by
\begin{equation}
    \varrho c \left[\frac{\partial T^*}{\partial t}+ \boldsymbol{v}\cdot\boldsymbol{\nabla} T^*\right]=\boldsymbol{\nabla}(k \boldsymbol{\nabla} T^*).
\end{equation}

Here, the thermal conductivity and the volumetric heat capacity of water are denoted by $k$ and $\varrho c$, respectively.

The equation for the conservation of cells can be written as follows \cite{ref1,ref9}:
\begin{eqnarray}
    \label{eqn:equation10}
    \frac{\partial n}{\partial t}=-\boldsymbol{\nabla}\cdot\boldsymbol{A}, \nonumber\\ 
    \boldsymbol{A}=n \boldsymbol{U}_c+n\boldsymbol{v}-\boldsymbol{D}\cdot\boldsymbol{\nabla} n,
\end{eqnarray}
where $\boldsymbol{A}$ represents the flux of cells and is defined as the sum of three components: the mean cells' swimming velocity $\boldsymbol{U}_c$, the flux caused by the advection of the cells, and the random component of the cell's locomotion. $\boldsymbol{D}=D\boldsymbol{I}$ depicts the isotropic and constant diffusivity tensor where $\boldsymbol{I}$ representing the identity tensor and $D$ is the diffusion coefficient.

The mean cell swimming velocity is given by
\begin{equation}
\label{eqn:equation5}
    \boldsymbol{U}_c=U_c\bar{\boldsymbol{P}},
\end{equation}
where $U_c$ is the average cell swimming speed and $\bar{\boldsymbol{P}}$ is the average cell swimming direction which is calculated as \cite{ref9}
\begin{equation}
\label{eqn:equation6}
    \bar{\boldsymbol{P}}=T(\mathcal{G})\boldsymbol{k}.
\end{equation}

The formulation of taxis function $T(\mathcal{G})$ is defined as follows:
\begin{equation}
\label{eqn:equation7}
 T(\mathcal{G})  \left\{ \begin{array}{lll}
 \ge 0 & \mbox{for $\mathcal{G}_c \ge \mathcal{G}$};\\
         < 0 & \mbox{for $\mathcal{G}_c < \mathcal{G}$},
        \end{array} \right. 
\end{equation}
The precise expression of the phototaxis function $T(\mathcal{G})$, varies among microorganisms due to their distinctive characteristics \cite{ref9}. To provide a further level of understanding of this concept, a particular implementation of the phototaxis function might be described as follows: \cite{ref13,ref17}
\begin{equation*}
    T(\mathcal{G})=0.8\sin{\left(\frac{3}{2}\pi\varphi(\mathcal{G})\right)}-0.1\sin{\left(\frac{1}{2}\pi\varphi(\mathcal{G})\right)},
\end{equation*}
where  $\varphi=\mathcal{G}e^{\chi(\mathcal{G}-1)}$. The parameter  $\chi$ depends on the critical total intensity $\mathcal{G}_c$.

Maintaining mass and cell conservation in an incompressible fluid suspension requires horizontal boundaries. No fluid or cell movement across these boundaries is required to preserve these concepts. The vertical velocity and cell movement at these borders must be zero. The fluid velocity at these boundaries depends on the type of surface. One makes a distinction between a stress-free boundary and a rigid one. Tangential stress must be zero for a stress-free boundary. This stress-free surface permits the fluid to flow parallel to the boundary. Rigid boundaries have a distinct impact, it inhibits fluid flow. This requires all velocity components to be zero. The fluid can't move along a rigid boundary, therefore its velocity in all directions must disappear.
The boundary conditions for the stress-free top surface
\begin{align}
\label{eqn:equation12}
    \boldsymbol{v}\cdot\boldsymbol{k}=0, \quad \frac{\partial^2}{\partial x_3^2}(\boldsymbol{v}\cdot\boldsymbol{k})=0, \quad T^*&=T_u^*, \quad \boldsymbol{A}\cdot\boldsymbol{k}=0,  \quad \text{at} \quad x_3=H,
\end{align}
while for the rigid bottom surface 
\begin{equation}
\label{eqn:equation11}
    \boldsymbol{v}=0, \quad \boldsymbol{v}\times\boldsymbol{k}=0, \quad  T^*=T_l^*, \quad\boldsymbol{A}\cdot\boldsymbol{k}=0,\quad \text{at} \quad x_3=0.
\end{equation}

The radiation transfer equation \cite{ref-modest,ref-chand}
\begin{equation}
\label{eqn:equation3}
    \boldsymbol{r}\cdot \boldsymbol{\nabla}I(\boldsymbol{x},\boldsymbol{r})+\psi I(\boldsymbol{x},\boldsymbol{r})=0,
\end{equation}
is used to calculate the intensity $I(\boldsymbol{x},\boldsymbol{r})$ of a light beam passing through an absorbing medium that does not scatter light. In this context, the absorption coefficient $\psi$ is defined with the concentration $n$ by a linear proportionality, which can be written as $\psi=\iota n$, where $\iota$ is the proportionality constant.
The boundary conditions are such that light reflection on the top and bottom surfaces is not considered. Therefore, the top surface condition can be written as
\begin{equation}
\label{eqn:equation4}
    I(x_1,x_2,H,\alpha_1,\alpha_2)=I^0\delta(\boldsymbol{r}-\boldsymbol{r}^0) , \quad \pi/2\le\alpha_1\le\pi,
\end{equation}
and the bottom surface condition as 
\begin{equation}
    I(x_1,x_2,0,\alpha_1,\alpha_2)=0,\quad 0\le\alpha_1\le\pi/2.
\end{equation}

The incident direction is denoted by $\boldsymbol{r}^0=-\boldsymbol{k}$, the incident irradiation is given by $I^0$, and the Dirac delta function is subject to the condition \cite{ref-modest}
\begin{equation*}
    \int_0^{4\pi} f(\boldsymbol{r})\delta(\boldsymbol{r}-\boldsymbol{r}^0) d\omega=f(\boldsymbol{r}^0).
\end{equation*}

At a point $\boldsymbol{x}=(x_1,x_2,x_3)$ the radiative heat flux $\boldsymbol{q}(\boldsymbol{x})$ and the total intensity $\mathcal{G}(\boldsymbol{x})$ are given as \cite{ref-modest}
\begin{equation}
\label{eqn:equation2}
    \boldsymbol{q}(\boldsymbol{x})=\int_0^{4\pi}I(\boldsymbol{x},\boldsymbol{r})\boldsymbol{r}d\omega,
\end{equation}
\begin{equation}
\label{eqn:equation1}
    \mathcal{G}(\boldsymbol{x})=\int_0^{4\pi}I(\boldsymbol{x},\boldsymbol{r})d\omega.
\end{equation}
Here the solid angle is indicated by $\omega$.

To construct the non-dimensional bioconvection equations, one must first scale all of the necessary factors, including lengths, pressure, fluid velocity, time, and cell concentration. This scaling is accomplished by the use of standard parameters including $H$, $\mu \alpha_f/H^{2}$, $\alpha_f/H$, $H^{2}/\alpha_f$, and $\bar{n}$. For the temperature, the non-dimensionalization is accomplished by defining $\mathcal{T}=\frac{T^*-T_u^*}{\Delta T}$, where $\alpha_f$ stands for the thermal diffusivity of water and $\Delta T= T_l^*-T_h^*$, which is the temperature difference that occurs between the upper and lower borders.

After performing the required changes to the governing equations (\ref{eqn:equation8})-(\ref{eqn:equation10}) by replacing corresponding dimensionless variables:
\begin{equation}
\label{eqn:equation13}
    div(\boldsymbol{v})=0, 
\end{equation}
\begin{equation}
\label{eqn:equation14}
    \frac{1}{P_r}\left(\frac{\partial}{\partial t}+\boldsymbol{v}\cdot\boldsymbol{\nabla}\right)\boldsymbol{v}=\nabla^2\boldsymbol{v}-\boldsymbol{\nabla} \mathcal{P}-n R_a \boldsymbol{k}-R_m \boldsymbol{k}+R_T \mathcal{T} \boldsymbol{k},
\end{equation}
\begin{equation}
    \frac{\partial \mathcal{T} }{\partial t}+ \boldsymbol{v}\cdot \boldsymbol{\nabla} \mathcal{T}=\boldsymbol{\nabla}^2 \mathcal{T},
\end{equation}
\begin{equation}
    \label{eqn:equation15}
    \frac{\partial n}{\partial t}=-\boldsymbol{\nabla}\cdot\left(\frac{1}{Le}n U_s\bar{\boldsymbol{P}}+n\boldsymbol{v}-\frac{1}{Le}\boldsymbol{\nabla} n\right).
\end{equation}

Here, the basic-density Rayleigh number is $R_m=\frac{\varrho g H^3}{\mu \alpha_f}$, the bioconvection Rayleigh number is $R_a=\frac{\bar{n}\vartheta \Delta \rho g H^3}{\mu \alpha_f}$, the thermal Rayleigh number is $R_T=\frac{\beta \Delta T \varrho g H^3}{\mu \alpha_f}$, the Lewis number is $Le=\frac{\alpha_f}{D}$, the Prandtl number is $P_r=\frac{\mu}{\varrho \alpha_f}$, dimensionless swimming speed is $U_s=\frac{U_c H}{D}$, and the kinematic viscosity is $\nu=\mu/\rho$.

Non-dimensional boundary conditions for the stress-free top surface become
\begin{align}
\label{eqn:equation17}
    \boldsymbol{v}\cdot\boldsymbol{k}&=0, \quad \frac{\partial^2}{\partial x_3^2}(\boldsymbol{v}.\boldsymbol{k})=0, \quad \mathcal{T}=0, \quad \nonumber \\
    &\left(\frac{1}{Le}n U_s\bar{\boldsymbol{P}}+n\boldsymbol{v}-\frac{1}{Le}\boldsymbol{\nabla} n\right)\cdot\boldsymbol{k}=0, \quad  \text{at} \quad x_3=1,
\end{align}
while for the rigid bottom surface
\begin{align}
\label{eqn:equation16}
    \boldsymbol{v}&=0, \quad \boldsymbol{v}\times\boldsymbol{k}=0, \quad  \mathcal{T}=1, \quad \nonumber\\
    &\left(\frac{1}{Le}n U_s\bar{\boldsymbol{P}}+n\boldsymbol{v}-\frac{1}{Le}\boldsymbol{\nabla} n\right)\cdot\boldsymbol{k}=0, \quad \text{at} \quad x_3=0.
\end{align}

Non-dimensional Radiative transfer equation (\ref{eqn:equation3}) becomes
\begin{equation}
    \frac{d I}{d r}+n \hbar  I(\boldsymbol{x},\boldsymbol{r})=0,
\end{equation}
where the optical depth of the suspension is $\hbar=\iota\bar{n}H$. 
The non-dimensional boundary condition for the top surface intensity becomes
\begin{equation}
    I(x_1,x_2,1,\alpha_1,\alpha_2)=I^0\delta(\boldsymbol{r}-\boldsymbol{r}^0) , \quad \pi/2\le\alpha_1\le\pi,
\end{equation}
while for the bottom surface intensity
\begin{equation}
    I(x_1,x_2,0,\alpha_1,\alpha_2)=0,\quad 0\le\alpha_1\le\pi/2.
\end{equation}

\section{Steady-state}
\label{sec3}
In the basic state, $\boldsymbol{v}=0$, $\mathcal{P}=\mathcal{P}_b$, $n=n_b(x_3)$, $I=I_b(x_3,\alpha_1)$ and $\mathcal{G}=\mathcal{G}_b(x_3)$ $\mathcal{T}=\mathcal{T}_b$. 

The radiative transfer equation at steady-state becomes
\begin{equation}
\label{eqn:equation18}
    \frac{\partial I_b}{\partial x_3}+\frac{\hbar n_b(x_3)}{\cos{\alpha_1}}I_b(x_3,\alpha_1)=0.
\end{equation}
The corresponding boundary condition is 
\begin{equation}
\label{eqn:equation19}
    I_b(1,\alpha_1)=I^0\delta(\boldsymbol{r}-\boldsymbol{r}^0).
\end{equation}

After resolving equation (\ref{eqn:equation18}), we get
\begin{equation}
\label{eqn:equation20}
    I_b(x_3,\alpha_1)=C\exp{\left(\frac{-\hbar}{\cos{\alpha_1}}\int_1^{x_3} n_b(s)ds\right)},
\end{equation}
apply boundary condition (\ref{eqn:equation19}), 
\begin{equation}
\label{eqn:equation21}
    I_b(x_3,\alpha_1)=I^0\delta(\boldsymbol{r}-\boldsymbol{r}^0)\exp{\left(\frac{-\hbar}{\cos{\alpha_1}}\int_1^{x_3} n_b(s)ds\right)}.
\end{equation}
Basic state total intensity $\mathcal{G}$ becomes
\begin{align}
\label{eqn:equation22}
    \mathcal{G}_b(x_3)&=\int_0^{4\pi} I_b(x_3,\alpha_1)d\omega \nonumber\\
   & =I^0\exp{\left(\hbar\int_1^{x_3} n_b(s)ds\right)}.
\end{align}

The basic state cell conservation equation can be transformed into 
\begin{equation}
\label{eqn:equation24}
\frac{d n_b}{dx_3}-U_s T_b n_b=0,
\end{equation}
with 
\begin{equation}
\label{eqn:equation25}
    \int_0^1 n_b(x_3)dx_3=1.
\end{equation}

The basic state radiative heat flux is written as
\begin{align}
\label{eqn:equation26}
    \boldsymbol{q}_b(x_3)&=\int_0^{4\pi}I_b(x_3,\alpha_1)\boldsymbol{r}d\omega \nonumber \\
    &=-I^0\exp{\left(\hbar\int_1^{x_3} n_b(s)ds\right)\boldsymbol{k}} \nonumber\\
    &=|\boldsymbol{q}_b|(-\boldsymbol{k}).
\end{align}

Therefore, mean swimming orientation $\bar{\boldsymbol{P}}_b$ at the basic state is calculated via 
\begin{equation}
\label{eqn:equation27}
    \bar{\boldsymbol{P}}_b=-T(\mathcal{G}_b)\frac{\boldsymbol{q}_b}{|\boldsymbol{q}_b|}=T(\mathcal{G}_b)\boldsymbol{k}.
\end{equation}

The basic state temperature is given by
\begin{equation}
    \mathcal{T}_b=1-x_3.
\end{equation}

\section{Linear stability analysis}
\label{sec4}
To explore linear instability, an infinitesimal perturbation $\epsilon(0<\epsilon \ll 1)$ is applied in the basic state via

\begin{equation*}
    \boldsymbol{v}=\boldsymbol{0}+\epsilon \boldsymbol{v}^*(x_1,x_2,x_3,t)+O(\epsilon^2),\\
    \end{equation*}
    \begin{equation*}
    n=n_b(x_3)+\epsilon n^*(x_1,x_2,x_3,t)+O(\epsilon^2),\\
    \end{equation*}
    \begin{equation*}
    \mathcal{P}=\mathcal{P}_b+\epsilon \mathcal{P}^*+O(\epsilon^2),\\
    \end{equation*}
    \begin{equation*}
    \bar{\boldsymbol{P}}=\bar{\boldsymbol{P}}_b+\epsilon\bar{\boldsymbol{P}}^*+O(\epsilon^2),\\
    \end{equation*}
    \begin{equation*}
    \mathcal{G}=\mathcal{G}_b+\epsilon\mathcal{G^*}+O(\epsilon^2),\\
    \end{equation*}
    \begin{equation*}
    \mathcal{T}=\mathcal{T}_b+\epsilon T'+O(\epsilon^2),
    \end{equation*}

here perturbed fluid velocity is $\boldsymbol{v}^* = (u^* , v^* , w^*)$.
Thus, the linearized governing equations are written as
\begin{equation}
\label{eqn:equation28a}
    div(\boldsymbol{v}^*) =0,  
\end{equation}
\begin{equation}
\label{eqn:equation28b}
   \frac{1}{P_r}\frac{\partial \boldsymbol{v}^*}{\partial t}=\nabla^2\boldsymbol{v}^*-\boldsymbol{\nabla} \mathcal{P}^*-n^* R_a \boldsymbol{k}+R_T T' \boldsymbol{k},
\end{equation}
\begin{equation}
    \frac{\partial T'}{\partial t}-\boldsymbol{v}\cdot\boldsymbol{k}=\boldsymbol{\nabla}^2 T',
\end{equation}
\begin{equation}
\label{eqn:equation29}
    \frac{\partial n^*}{\partial t}=-\frac{dn_b}{dx_3}w^*+ \frac{1}{Le}\nabla^2 n^*-\frac{1}{Le}U_s\boldsymbol{\nabla}\cdot(n^*\bar{\boldsymbol{P}}_b+n_b\bar{\boldsymbol{P}}^*).
\end{equation}

 The total intensity can be written as
 \begin{align*}
     \mathcal{G}&=\mathcal{G}_b+\epsilon\mathcal{G^*}+O(\epsilon^2)\nonumber\\
     &=I^0\exp{\left(\hbar\int_1^{x_3}(n_b(s)+\epsilon n^*+O(\epsilon^2))ds\right)}.
 \end{align*}
On accumulating $O(\epsilon)$ terms, perturbed total intensity becomes
 \begin{equation*}
    \mathcal{G}^*=I^0\left(\hbar\int_1^{x_3} n^* ds\right)\exp{\left(\hbar\int_1^{x_3} n_b(s)ds\right)}.
\end{equation*}
The mean swimming orientation is given by
\begin{align}
\label{eqn:equation30}
    \bar{\boldsymbol{P}}&=\bar{\boldsymbol{P}}_b+\epsilon\bar{\boldsymbol{P}}^* +O(\epsilon^2) \nonumber \\
    &=T(\mathcal{G}_b+\epsilon \mathcal{G}^*+O(\epsilon^2))\boldsymbol{k}.
\end{align}
On accumulating $O(\epsilon)$ terms, perturbed mean swimming orientation becomes
\begin{equation}
\label{eqn:equation31}
    \bar{\boldsymbol{P}}^*=\mathcal{G}^*\frac{\partial T}{\partial \mathcal{G}_b}\boldsymbol{k}.
\end{equation}

Applying the curl operator twice to (\ref{eqn:equation28b}) and focusing on the vertical component yields 
\begin{align}
\label{eqn:equation34}
    \frac{1}{P_r}\frac{\partial}{\partial t}(\boldsymbol{\nabla}^2 w^*)&=\boldsymbol{\nabla}^4 w^*-R_a \boldsymbol{\nabla}^2 n^*+R_a\frac{\partial^2}{\partial x_3^2} n^*
    -R_T \boldsymbol{\nabla}^2 T'+R_T\frac{\partial^2}{\partial x_3^2} T'.
\end{align}

In equation (\ref{eqn:equation29})
\begin{eqnarray}
\label{eqn:equation32}
    \boldsymbol{\nabla}\cdot(n^*\bar{\boldsymbol{P}}_b+n_b\bar{\boldsymbol{P}}^*)=2\hbar n_b \mathcal{G}_b\frac{d T(\mathcal{G}_b)}{d \mathcal{G}_b}+T(\mathcal{G}_b)\frac{\partial n^*}{\partial x_3}
    +\hbar\frac{\partial}{\partial x_3}\left(n_b\mathcal{G}_b\frac{d T(\mathcal{G}_b)}{d\mathcal{G}_b}\right)\int_1^{x_3} n^* ds.
\end{eqnarray}
Therefore, equation (\ref{eqn:equation29}) re-write as
\begin{align}
\label{eqn:equation33}
    &\frac{\partial n^*}{\partial t}-\frac{1}{Le}\boldsymbol{\nabla}^2 n^*-\frac{1}{Le}\hbar U_s\frac{\partial}{\partial x_3}\left(n_b\mathcal{G}_b\frac{d T(\mathcal{G}_b)}{d\mathcal{G}_b}\right)\int_{x_3}^1  n^* ds \nonumber\\
    &+2\frac{1}{Le}\hbar n_b \mathcal{G}_b\frac{d T(\mathcal{G}_b)}{d \mathcal{G}_b}n^*U_s+\frac{1}{Le}T(\mathcal{G}_b)\frac{\partial n^*}{\partial x_3}U_s
    =-w^*\frac{d n_b}{d x_3}.
\end{align}

Perturbed boundary conditions for the stress-free top surface become
\begin{align*}
    w^*=0, \quad \frac{\partial^2 w^*}{\partial {x_3}^2}=0, \quad T'&=0,\quad U_s T_b n^*-\frac{\partial n^*}{\partial x_3}=0, \text{at} \quad x_3=1,
\end{align*}
while for the rigid bottom surface
\begin{align*}
    w^*=0, \quad \frac{\partial w^*}{\partial x_3}=0, \quad T'=0, \quad 
    \hbar U_s n_b \mathcal{G}_b \frac{d T_b}{d \mathcal{G}_b} \int_{x_3}^1 n^* d \bar{x}_3-U_s T_b n^*+\frac{\partial n^*}{\partial x_3}=0 \quad 
    \text{at} \quad x_3=0.
\end{align*}

Normal modes are decomposed from linearized governing equations via 
\begin{equation*}
    w^*=\hat{w}(x_3)\exp{[\gamma t+i(a_1 x_1+a_2 x_2)]},
\end{equation*}
\begin{equation*}
    n^*=\hat{n}(x_3)\exp{[\gamma t+i(a_1 x_1+a_2 x_2)]},
\end{equation*}
\begin{equation*}
   T'=\Theta(x_3)\exp{[\gamma t+i(a_1 x_1+a_2 x_2)]}.
\end{equation*}
Here, $a_1$ and $a_2$ are wavenumbers in the $x_1$ and $x_2$ directions and the resultant $a=\sqrt{a_1^2+a_2^2}$ is a horizontal wavenumber.

Thus, the governing equations (\ref{eqn:equation28a})-(\ref{eqn:equation29}) in normal modes become 
\begin{align}
    \label{eqn:equation36}
    \frac{\gamma}{P_r}\left(\frac{d^2}{dx_3^2}-a^2\right) \hat{w}(x_3) -\left(\frac{d^2}{dx_3^2}-a^2\right)^2\hat{w}(x_3) 
    = a^2R_a\hat{n}(x_3)-a^2R_T\Theta(x_3),
\end{align}
\begin{equation}
    \label{eqn:equation37}
    \left(\gamma+a^2-\frac{d^2}{d x_3^2}\right)\Theta=\hat{w},
\end{equation}
\begin{eqnarray}
\label{eqn:equation38}
    -\hbar U_s \frac{\partial}{\partial x_3}\left(n_b\mathcal{G}_b\frac{d T(\mathcal{G}_b)}{d\mathcal{G}_b}\right)\int_{x_3}^1  \hat{n}d \bar{x}_3+( \gamma Le+a^2) \hat{n}+&2\hbar n_b \mathcal{G}_b\frac{d T_b}{d \mathcal{G}_b}U_s \hat{n}+U_s T_b \frac{d \hat{n}}{d x_3}\nonumber\\
    &-\frac{d^2 \hat{n}}{d x_3^2}
    =-Le \frac{d n_b}{d x_3}\hat{w}.
\end{eqnarray}

The associated boundary conditions in the normal modes for the stress-free top surface become
\begin{align*}
    \hat{w}(x_3)=0,\quad \frac{d^2\hat{w}(x_3)}{d x_3^2}=0,\quad \Theta=0, 
    \quad U_s T_b \hat{n}-\frac{d \hat{n}}{d x_3}=0  \quad \text{at} \quad x_3=1,
\end{align*}
while for the rigid bottom surface
\begin{align*}
    \hat{w}(x_3)=0,\quad\frac{\hbar\hat{w}(x_3)}{d x_3}=0,\quad \Theta=0,\quad 
    \hbar U_s n_b \mathcal{G}_b \frac{d T_b}{d \mathcal{G}_b} \int_{x_3}^1 \hat{n} d \bar{x}_3-U_s T_b \hat{n}+\frac{d \hat{n}}{d x_3}=0\quad 
    \text{at} \quad x_3=0.
\end{align*}

\begin{figure*}[!htbp]
    \centering
    \includegraphics[width=15cm, height=15cm ]{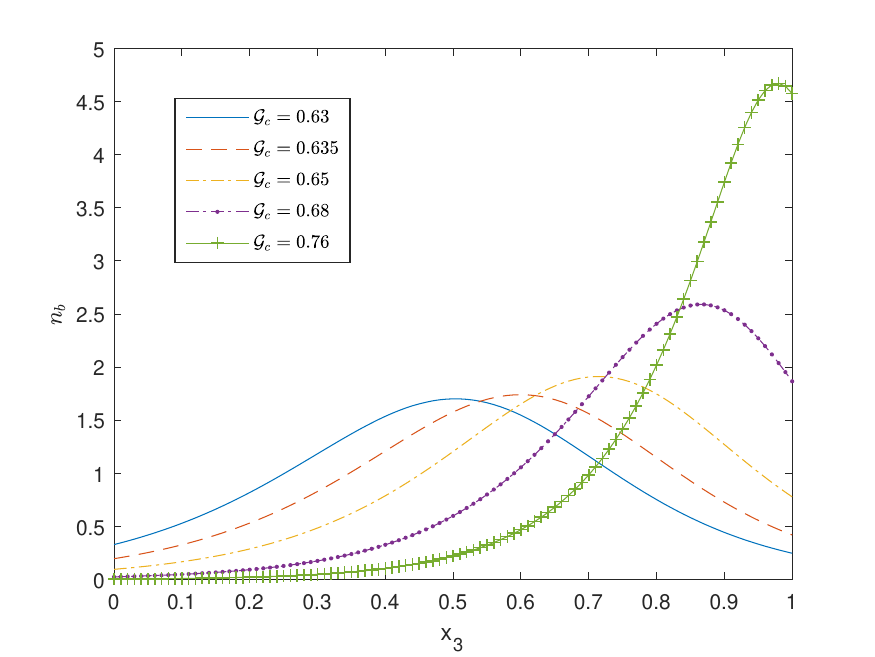}
    \caption{Effect of critical total intensity $\mathcal{G}_c$ on the steady-state concentration profile for fixed parameters $U_s=10$, and $\hbar=0.5$.  }
   \label{fig:intensity.pdf}
 \end{figure*}
\begin{figure*}[!htbp]
    \centering
    \includegraphics[width=15cm, height=15cm ]{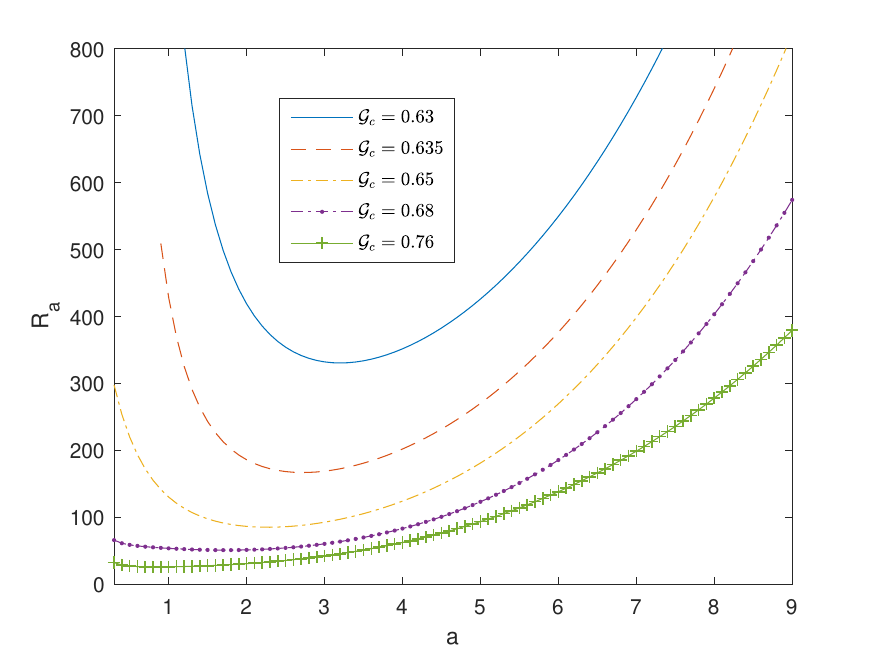}
    \caption{Neutral curves fixed parameters $\hbar=0.5$, $U_s=10$, $R_T=50$, and $Le=4$ as $\mathcal{G}_c$ is varied.  }
   \label{fig:10v0.5k50Rt4Lestn.pdf}
 \end{figure*}
\begin{figure*}[!htbp]
    \centering
    \includegraphics[width=15cm, height=15cm ]{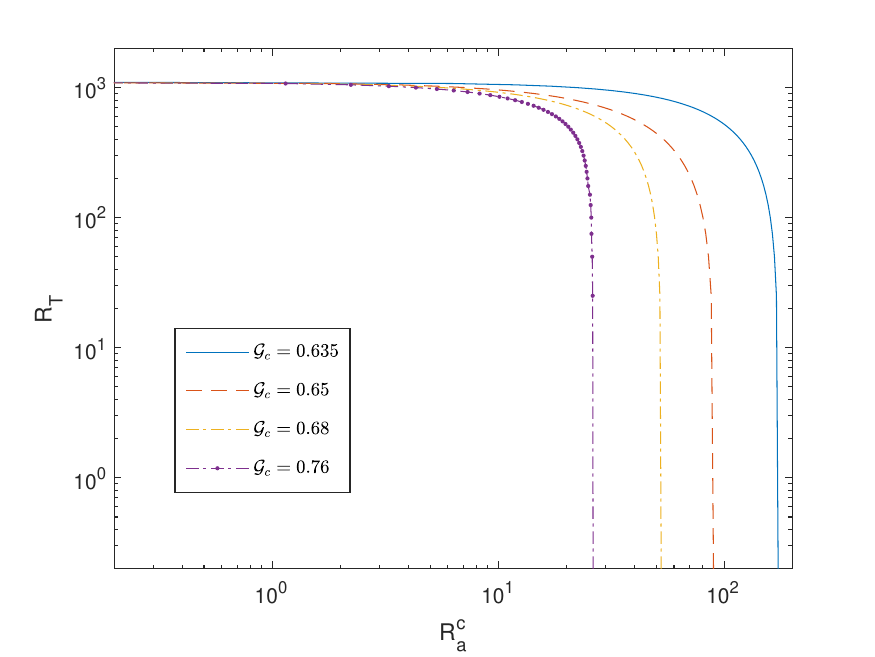}
    \caption{The relationship between $R_a^c$ and $R_T$ for $\hbar=0.5$, $U_s=10$, and $Le=4$ as $\mathcal{G}_c$ is varied. }
   \label{fig:figbasicstab.pdf}
 \end{figure*}
\begin{figure*}[!htbp]
    \centering
    \includegraphics[width=15cm, height=15cm ]{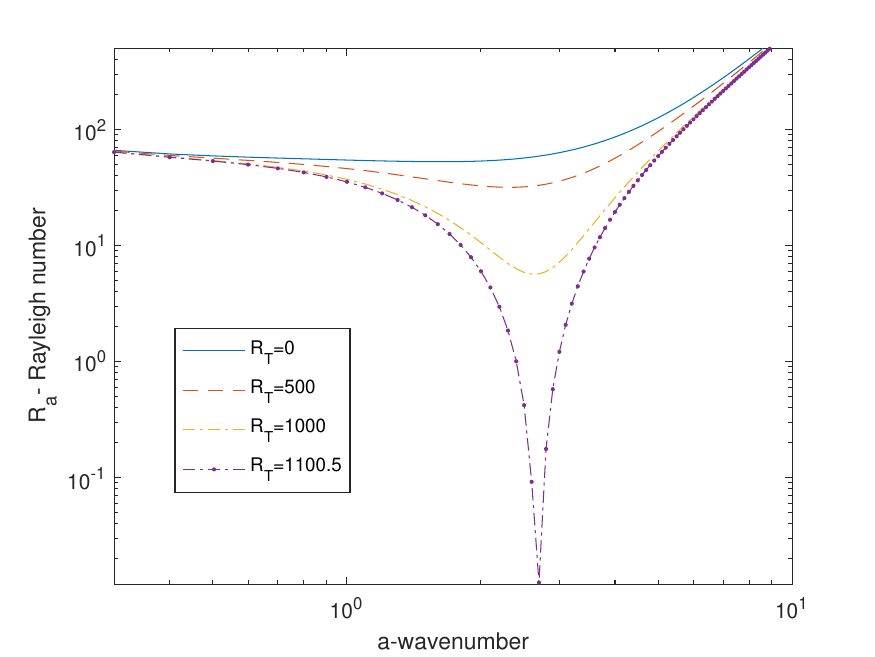}
    \caption{Neutral curves for $\hbar=0.5$, $U_s=10$, $\chi=0.19$, $\mathcal{G}_c=0.68$, $Le=4$ as $R_T$ is varied. }
   \label{fig:10v0.5k0.68Ic.pdf}
 \end{figure*}

Define a new variable
\begin{equation}
\label{eqn:equation39}
    N(x_3)=\int_{x_3}^1 \hat{n} d \bar{x}_3,
\end{equation}
so that the system of equations becomes
\begin{align}
\label{eqn:equation40}
    \frac{d^4 \hat{w}}{d x_3}-\left(2 a^2+ \frac{\gamma}{P_r}\right)\frac{d^2 \hat{w}}{d x_3^2}+a^2\left(a^2+\frac{\gamma}{P_r}\right)\hat{w}
    =a^2 R_a \frac{d N}{d x_3}+a^2 R_T \Theta,
\end{align}
\begin{equation}
\label{eqn:equation41}
    \left(\gamma+a^2-\frac{d^2}{d x_3^2}\right)\Theta=\hat{w},
\end{equation}
\begin{align}
\label{eqn:equation42}
    &\frac{d^3 N}{d x_3^3}-U_s T_b \frac{d^2 N}{d x_3^2}-\left(\gamma Le+a^2+2\hbar U_s n_b \mathcal{G}_b\frac{d T_b}{d \mathcal{G}_b}\right)\frac{d N}{d x_3}- \hbar U_s \frac{d}{d x_3}\left(n_b \mathcal{G}_b\frac{d T_b}{d \mathcal{G}_b}\right)N=-Le\frac{d n_b}{d x_3}\hat{w}.
\end{align}
Also, boundary conditions for the stress-free top surface become
\begin{align}
    \label{eqn:equation45}
    \hat{w}(x_3)=0,\quad \frac{d^2\hat{w}(x_3)}{d z^2}=0,\quad \Theta=0, \quad 
    U_s T_b \frac{d N}{d x_3}-\frac{d^2 N}{d x_3^2}=0 \quad  \text{at} \quad x_3=1,
\end{align}
while for the rigid bottom surface
\begin{align}
    \label{eqn:equation43}
    \hat{w}(x_3)=0,\quad \frac{d\hat{w}(x_3)}{d x_3}=0, \quad \Theta=0, \quad
    \hbar U_s n_b \mathcal{G}_b \frac{d T_b}{d \mathcal{G}_b} N+U_s T_b \frac{d N}{d x_3}
    -\frac{d^2 N}{d x_3^2}=0\quad  \text{at} \quad x_3=0,
\end{align}

and 
\begin{equation}
\label{eqn:equation46}
    N(x_3)=0 \quad \text{at} \quad x_3=1.
\end{equation}

A set of governing equations and boundary conditions (\ref{eqn:equation40})-(\ref{eqn:equation46}) constitute a system of phototactic thermal-bioconvection equations. Without thermal convection, the governing equations return to the form presented by Vincent and Hill \cite{ref9}.

\begin{table*}
\caption{Common suspension parameters for the phototactic microorganism \textit{Chlamydomonas}\cite{ref9,ref13,ref17,kumar2023effect,zhao2018linear}.}
\begin{tabular}{ p{8cm} p{5cm}}
\hline
\hline
Scaled average swimming speed &$U_s=20H$\\
Kinematic viscosity &$\nu=10^{-2}$cm$^2$/s\\
Prandtl number &$P_r=5$ \\
Average concentration &$\bar{n}=10^6$cm$^{-3}$\\
Average cell swimming speed &$U_c =10^{-2}$cm/s\\
Cell volume &$\vartheta=5\times10^{-10}$cm$^3$\\
Cell diffusivity &$D=5\times10^{-5}-5\times10^{-4}$cm$^2$/s\\
Thermal diffusivity&$\alpha_f=2\times 10^{-3}$cm$^2$/s\\
Volumetric thermal expansion coefficient&$\beta=3.4\times 10^{-3}$K$^{-1}$\\
Temperature of upper wall&$T_u^*=300$ K\\
Temperature difference& $\Delta T=1$K\\
Ratio of cell density &$\Delta \varrho/\varrho=5\times 10^{-2}$ \\
Cell radius &$10^{-3}$cm\\
\hline
\hline
\end{tabular}\\
\label{tab:table1}
\end{table*}

 \begin{table*}
\caption{\label{tab:table2} For various parameters $\hbar=0.5,1.0$, $U_s=10$, $R_T=0$, $Le=1$, the critical Rayleigh number is computed by Vincent and Hill\cite{ref9} and the present study.}
\begin{center}
\begin{tabular}{{p{1.75cm}p{1.75cm}p{1.75cm}p{1.75cm}p{2cm}p{2.75cm}p{1.75cm}}}
\hline\hline
$\hbar$& $U_s$ & $\mathcal{G}_c$&$\chi$&$R_a^c$ [Ref. \cite{ref9}]&$R_a^c$ (Present study)& Error ($\%$) \\
\hline
0.5 & 10 &0.7&0.29&165.85&170.42&2.75\\
0.5 & 10   & 0.68&0.19 &210.4&211.29 &0.42 \\
0.5 &10 & 0.667 &0.1&262.4 & 262.8 &0.15 \\
0.5   &10 & 0.66&0.078 & 282&282.25& 0.08\\
0.5 & 10 & 0.65&0.029 & 360.01& 360.21&0.05\\
0.5 &10& 0.635 &-0.03& 695.13&695.23 & 0.01 \\
0.5 & 10& 0.63 &-0.068& 1349.34& 1349.52&0.01 \\
1.0&10&0.55&-0.35&343.51&344.07&0.163\\
1.0&10&0.539&-0.39&368.81&369.52&0.19\\
1.0&10&0.535&-0.4&373.77&374.18&0.1\\
1.0&10&0.525&-0.43&386.82&387.07&0.06\\
1.0&10&0.5&-0.5&535.32&535.39&0.01\\
1.0&10&0.495&-0.523&877.31&877.43&0.01\\
\hline\hline
\end{tabular}
\end{center}
\end{table*}

\begin{figure*}[!htbp]
    \centering
    \includegraphics[width=15cm, height=15cm ]{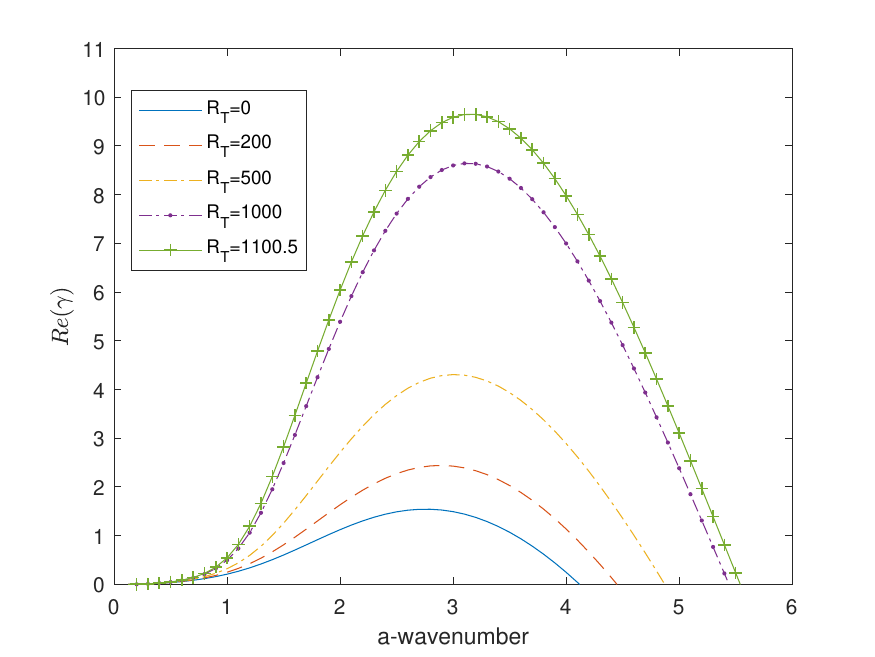}
    \caption{Growth rate for $\hbar=0.5$, $U_s=10$, $\chi=0.19$, $\mathcal{G}_c=0.68$, and $R_a=100$ as $R_T$ is varied.}
   \label{fig:growthrate.pdf}
 \end{figure*}
\begin{figure*}[!htbp]
    \centering
    \includegraphics[width=15cm, height=15cm ]{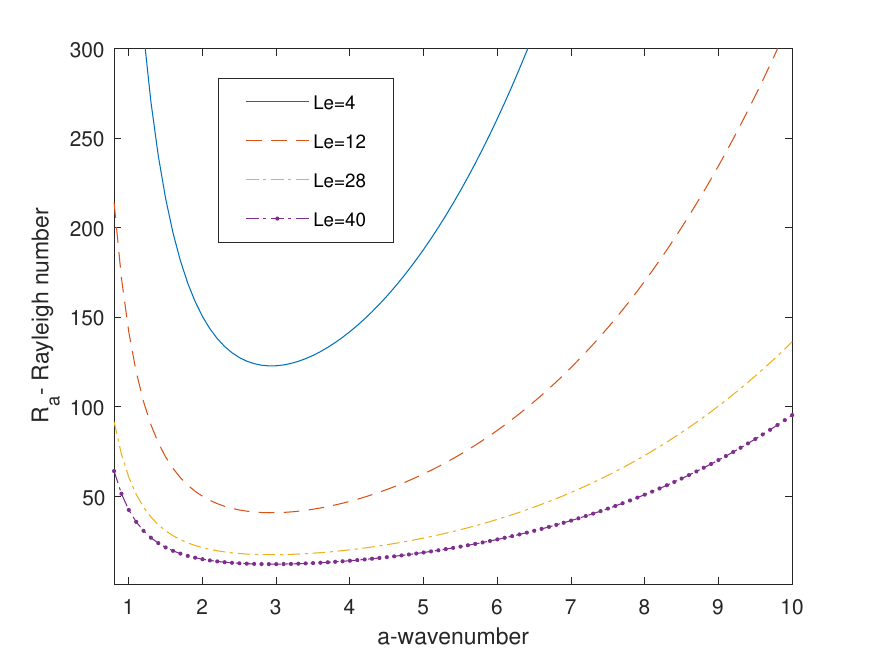}
    \caption{Neutral curves for $\hbar=1.0$, $U_s=10$, $\chi=-0.5$, and $\mathcal{G}_c=0.5$, and $R_T=100$ as $Le$ is varied.  }
   \label{fig:10v1k0.5Ic100Rt.pdf}
 \end{figure*}

 \begin{figure*}[!htbp]
    \centering
    \includegraphics[width=16cm, height=11cm ]{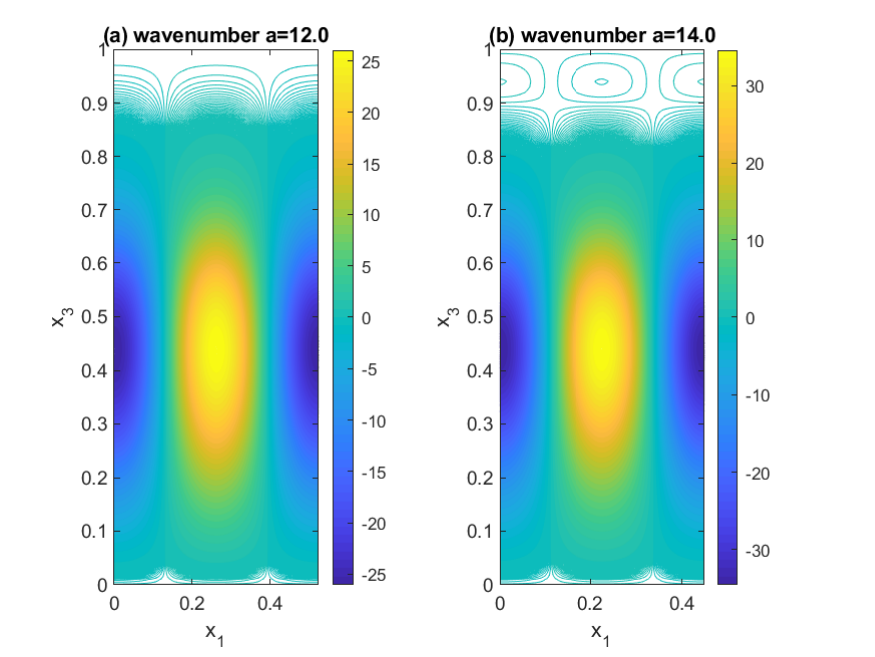}
    \caption{ Flow patterns on the stationary branch for $\hbar=1.0$, $U_s=10$, $\chi=-0.5$, and $\mathcal{G}_c=0.5$, $R_T=100$, $Le=4$, (a) $a=12.0$, and (b) $a=14.0$. }
   \label{fig:phaseportrait.pdf}
 \end{figure*}
  \begin{figure*}[!htbp]
    \centering
    \includegraphics[width=16cm, height=11cm ]{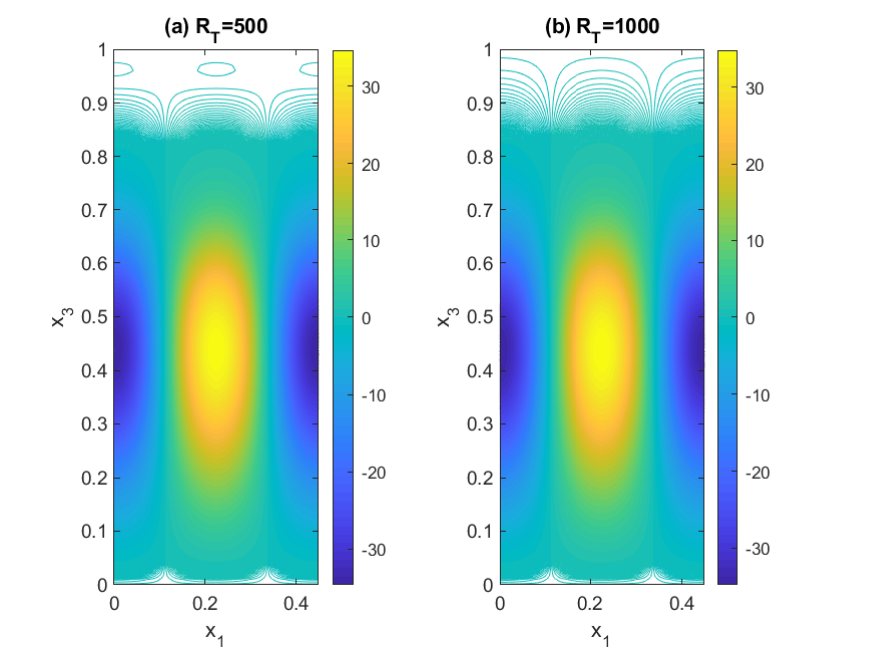}
    \caption{ Flow patterns on the stationary branch for $\hbar=1.0$, $U_s=10$, $\chi=-0.5$, and $\mathcal{G}_c=0.5$, $a=14.0$, $Le=4$, (a) $R_T=500$, and (b) $R_T=1000$. }
   \label{fig:phaseportraitRt.pdf}
 \end{figure*}
\begin{figure*}[!htbp]
    \centering
    \includegraphics[width=15cm, height=15cm ]{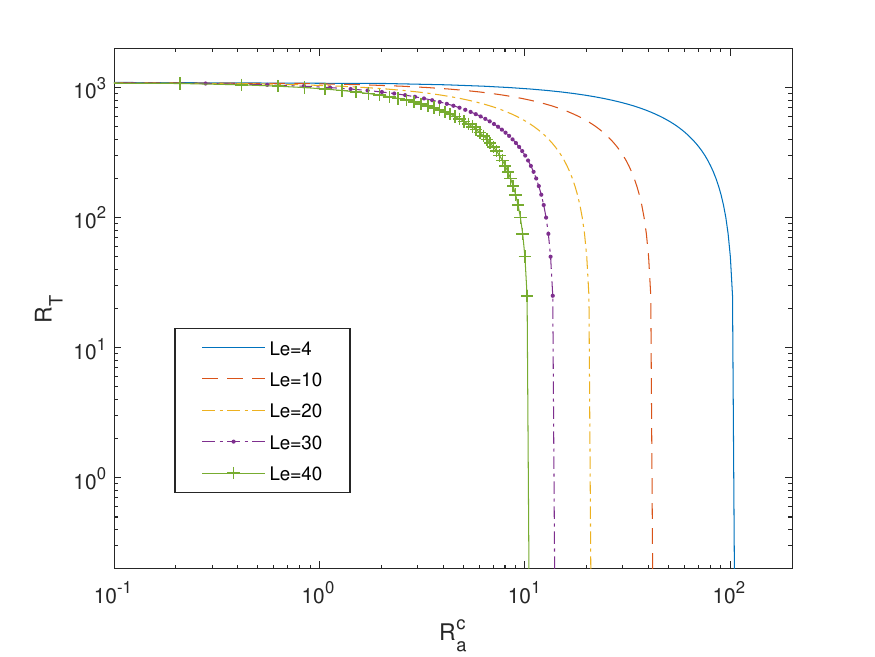}
    \caption{The relationship between $R_a^c$ and $R_T$ for $\hbar=0.5$, $U_s=15$, $\chi=-0.02$, and $\mathcal{G}_c=0.64$ as $Le$ is varied. }
   \label{fig:15v0.5klewis.pdf}
 \end{figure*}
   
 \begin{figure*}[!htbp]
    \centering
    \includegraphics[width=15cm, height=15cm ]{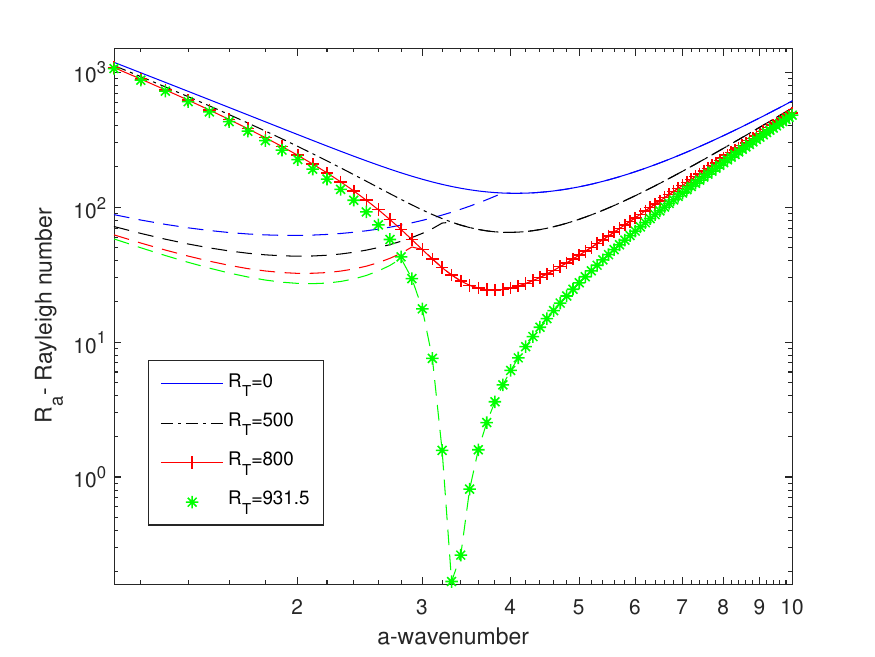}
    \caption{Neutral curves for $\hbar=1.0$, $U_s=15$, $\chi=-0.485$, $\mathcal{G}_c=0.51$, $Le=4$ as $R_T$ is varied. }
   \label{fig:15v1k.pdf}
 \end{figure*}

\begin{figure*}[!htbp]
    \centering
    \includegraphics[width=16cm, height=20cm ]{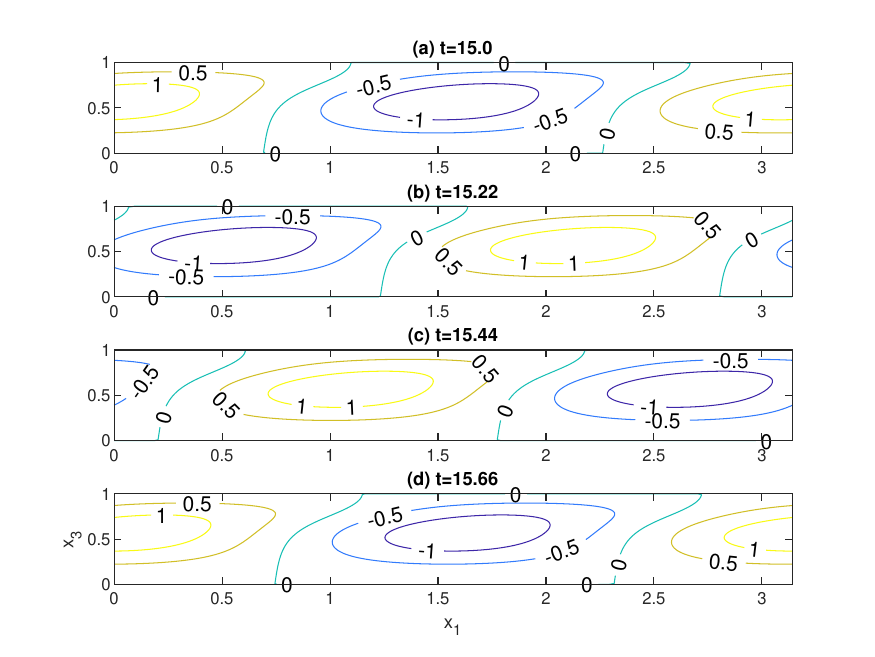}
    \caption{Flow pattern produced by the perturbed velocity $w^*$	 throughout one cycle of oscillation for  $\hbar=1.0$, $U_s=15$, $\chi=-0.485$, $\mathcal{G}_c=0.51$, $Le=4$, $R_T=500$, $R_a^c=43.38$, $a_c=2$, $Im(\gamma)=9.36$, (a) $t=15.0$, (b) $t=15.22$, (c) $t=15.44$, and (d) $t=15.66$.}
   \label{fig:periodic.pdf}
 \end{figure*}
 \begin{figure*}[!htbp]
    \centering
    \includegraphics[width=16cm, height=20cm ]{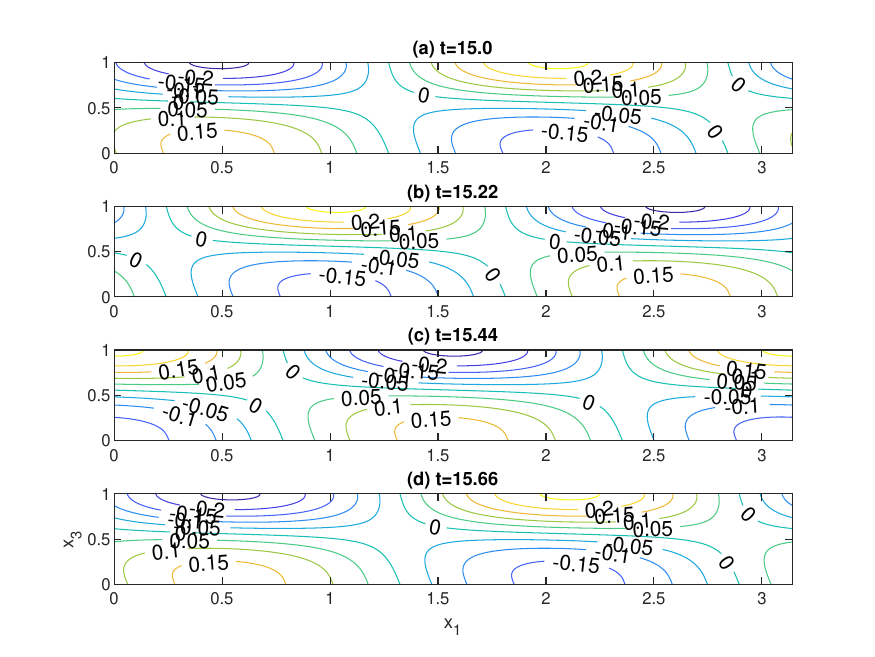}
    \caption{Flow pattern produced by the perturbed temperature $T'$ throughout one cycle of oscillation for  $\hbar=1.0$, $U_s=15$, $\chi=-0.485$, $\mathcal{G}_c=0.51$, $Le=4$, $R_T=500$, $R_a^c=43.38$, $a_c=2$, $Im(\gamma)=9.36$, (a) $t=15.0$, (b) $t=15.22$, (c) $t=15.44$, and (d) $t=15.66$.}
   \label{fig:periodicRt.pdf}
 \end{figure*}
\begin{figure*}[!htbp]
    \centering
    \includegraphics[width=16cm, height=11cm ]{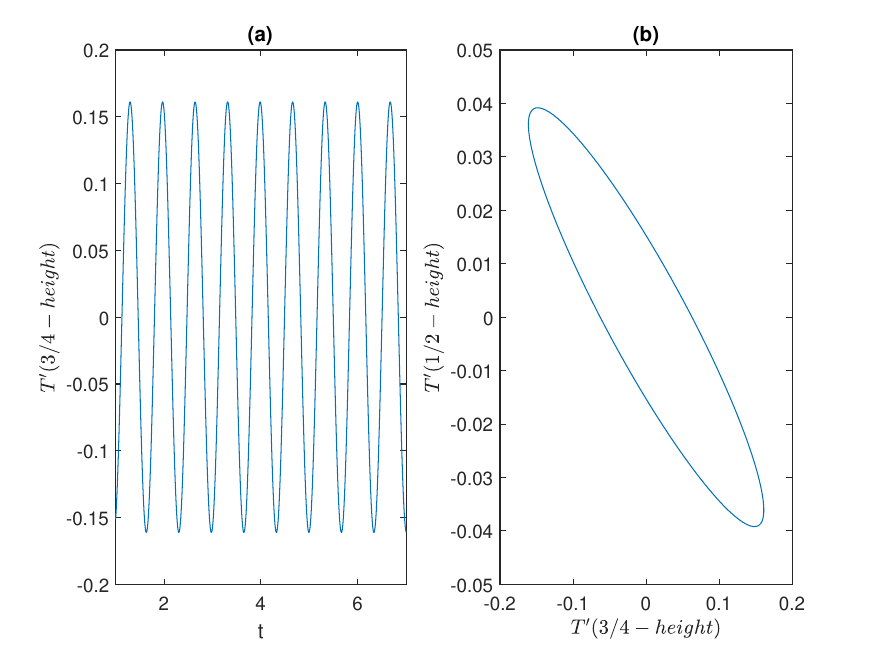}
    \caption{(a) Time-evolving perturbed temperature $T'$ and (b) phase diagram for $\hbar=1.0$, $U_s=15$, $\chi=-0.485$, $\mathcal{G}_c=0.51$, $Le=4$, $R_T=500$, $R_a^c=43.38$, $a_c=2$, $Im(\gamma)=9.36$}
   \label{fig:timeinv+limitcycleRt.pdf}
 \end{figure*}

\begin{figure*}[!htbp]
    \centering
    \includegraphics[width=15cm, height=15cm ]{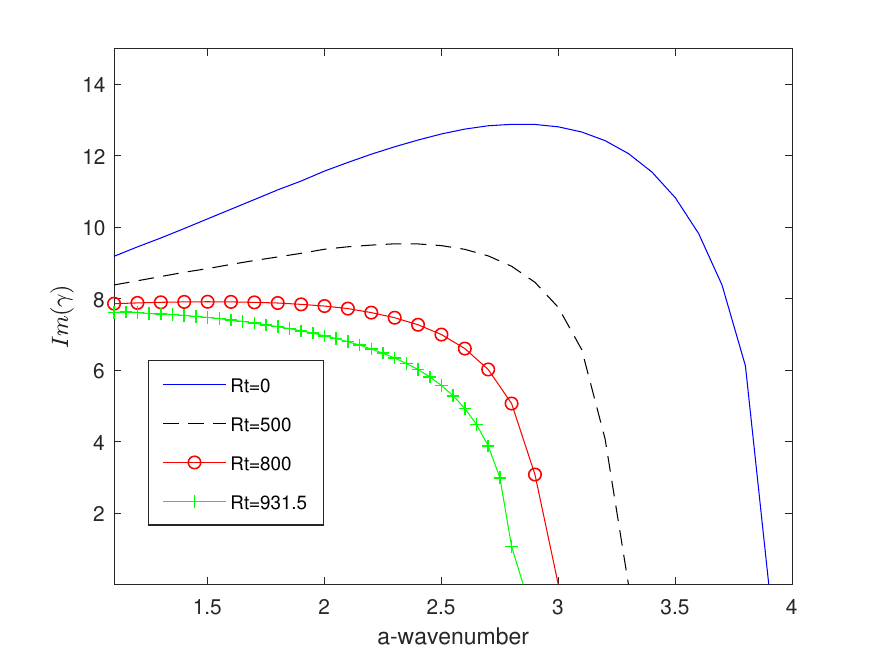}
    \caption{ Positive frequency curves for $\hbar=1.0$, $U_s=15$, $\chi=-0.485$, $\mathcal{G}_c=0.51$, and $Le=4$ as $R_T$ is varied. }
   \label{fig:frequency.pdf}
 \end{figure*}
 \section{Numerical results}
\label{sec5}
The set of ordinary differential equations (ODEs) represented by equations (\ref{eqn:equation40}) to (\ref{eqn:equation42}) constitutes a ninth-order system that is accompanied by nine boundary conditions (\ref{eqn:equation45}) to (\ref{eqn:equation46}). For dealing with coupled ODEs, a numerical method known as the shooting technique is used as part of the analysis approach. The MATLAB built-in bvp4c solver facilitates the computation \cite{shampine2003solving}. The investigation of the linear stability of the basic state will be the primary emphasis of this work. To do this, we will be illustrating neutral curves. These neutral curves are constituted by the points where the real component of the growth rate $Re(\gamma)$ is equal to zero. These zero points have relevance while evaluated in the context of stability analysis. The existence of an overstable or oscillatory solution is possible in cases if the imaginary component of the growth rate $Im(\gamma)$ along the neutral curve is not zero. On the other hand, according to the concept of exchange of stabilities, perturbations to the basic state are considered to be stationary if the value of $Im(\gamma)$ is equal to zero along this curve \cite{ref12}. In the time dependence which is expressed as $\exp(\gamma t) = \exp[(Re(\gamma)+i Im(\gamma))t]$, we include both the growth rate $Re(\gamma)$ as well as a hypothetical oscillation that is described by $Im(\gamma)/2\pi$. In situations with stable layering, the wave vectors have a negative value for the $Re(\gamma)$ parameter. However, as wave numbers venture just above the convective threshold, a narrow range comes into play where $Re(\gamma)$ becomes positive. The branch of the solution that exhibits the most significant amount of instability is the one in which the bioconvection Rayleigh number $R_a$ takes on its lowest possible value, which is represented by the symbol $R_a^c$. This critical solution, which has been given the name $(a^c, R_a^c)$, is known for being the solution deemed the most unstable. To bring our model into line with the work that has been done in the past on phototactic bioconvection, we take into account phototactic microorganisms which are analogous to \textit{Chlamydomonas}. Consequently, we establish the required parameters for this study in reference to existing works like Vincent and Hill \cite{ref9}, Ghorai and
Hill \cite{ref13}, Panda et al. \cite{ref17}, Kumar \cite{kumar2023effect} (refer to Table \ref{tab:table1}). The radiation parameters are the same as in Ref.\cite{ref14}. The optical depth can vary ranging from $0.25$ to $1$ for a suspension depth of $0.5$ cm, and $U_s = 10$ is the scaled swimming speed that corresponds to this range. Similarly, the value of $U_s$ for a suspension with a depth of $1.0$ cm is $20$. To maintain compatibility with several other phototactic bioconvection models, $I^0 = 0.8$ will be kept constant throughout. It has been demonstrated that there is a connection between the critical intensity $\mathcal{G}_c$ and the parameter in such a way that $-1.1\le\chi\le 1.1$ results in $0.3\le\mathcal{G}_c\le0.8$ \cite{ref17,kumar2023effect}.

For the neutral curves of linear stability with specific fixed parameters, Table (\ref{tab:table2}) demonstrates a perfect agreement between the Newton-Raphson-Kantorovich method \cite{ref9} and the bvp$4$c MATLAB technique. The relative error is seen to be less than $2.75\%$. Fig. (\ref{fig:intensity.pdf}) illustrates the impacts of critical intensity $\mathcal{G}_c$ on the basic concentration profile for fixed parameters $U_s=10$, and $\hbar=0.5$. For $\mathcal{G}_c=0.63$, the position of maximum concentration is at the middle of the suspension. At the sublayer position, the concentration of the microorganisms is maximum, whose location is determined by the critical total intensity value. The sublayer position shifts to the top of the suspension as the critical total intensity increases. Consequently, the position of maximum concentration shifts towards the top of the suspension. The value of the maximum concentration becomes highest near the top of the suspension, and the concentration gradient also increases. The region below the sublayer which is gravitationally unstable and supports the convection increases as critical intensity increases. Therefore, the corresponding critical Rayleigh number decreases and the system becomes more unstable (see Fig. \ref{fig:10v0.5k50Rt4Lestn.pdf}). The correlation between $R_a^c$ and $R_T$ is shown in Fig. (\ref{fig:figbasicstab.pdf}) for a variety of various values of $\mathcal{G}_c$. It has been shown that the $R_a^c$ reduces as the $R_T$ increases. This indicates that the rising temperature gradient throughout the suspension encourages the formation of bioconvection and results in the suspension being more unstable.

The impact of thermal Rayleigh number $R_T$ is described in Fig. (\ref{fig:10v0.5k0.68Ic.pdf}) for fixed parameters  $\hbar=0.5$, $U_s=10$, $\chi=0.19$, $\mathcal{G}_c=0.68$, $Le=4$. For $R_T=0$, the critical wavenumber $a^c$ is $1.6$, and the critical Rayleigh number $R_a^c$ is $52.78$. The principal effect of the thermal Rayleigh number can be seen in the neutral curves near the critical wavenumber. As the critical thermal Rayleigh number increases to $1100.5$, the critical Rayleigh number tends to zero. By illustrating growth rate curves for Rayleigh number values above $R_a^c$ $(R_a > R_a^c)$, it is possible to calculate approximations of the anticipated wavelengths of the initial perturbations. Fig. (\ref{fig:growthrate.pdf}) depicted growth rate for $\hbar=0.5$, $U_s=10$, $\chi=0.19$, $\mathcal{G}_c=0.68$, and $R_a=100$ as $R_T$ is varied. If the temperature $R_T$ rises, the critical Rayleigh number decreases, which causes the system to become more unstable. As a result, the growth rate that corresponds to it increases. In a pattern with a wavenumber of $a=2.8$ or a wavelength of $\lambda=2.24$, the value of the maximum growth rate is $Re(\gamma)_{max}=1.54$ for $R_T=0$. Maximum growth rates and their associated wavelengths both rise as $R_T$ is increased. For $Rt=1100.5$, the maximum growth rate is $9.65$ and the wavelength is $1.96$.

The impact of Lewis number $Le$ is described in Fig. (\ref{fig:10v1k0.5Ic100Rt.pdf}) for fixed parameters $\hbar=1.0$, $U_s=10$, $\chi=-0.5$, and $\mathcal{G}_c=0.5$, $R_T=100$. From Table (\ref{tab:table1}) it is apparent that $D$ can vary between $5\times 10 ^{-5} cm^2/s$ and $5\times 10 ^{-4} cm^2/s$, leading to the Lewis number variation between $4$ and $40$. The illustration demonstrates that a rise in Lewis number $Le$ results in a decrease in the critical Rayleigh number $R_a^c$. This indicates that $Le$ has a stabilizing effect on the system. For the parameters, $\hbar=1.0$, $U_s=10$, $\chi=-0.5$, and $\mathcal{G}_c=0.5$, $R_T=100$, $Le=4$, $a=12.0$, and $a=14.0$, the flow pattern of the perturbed velocity $w^*$ is displayed in Fig. (\ref{fig:phaseportrait.pdf}). If a solution has $n$ convection cells that are vertically layered one above the other, then that solution has mode $n$. In the case when $a=12.0$, the bioconvection solution consists of a single convection cell throughout the suspension. As a result, the bioconvective solution exhibits characteristics of the mode $1$ type at $a=12.0$. A zone of downwelling is indicated by a velocity with a value of negative, whereas an upwelling region is shown by a positive velocity. The solitary convection cells that formed layers along the suspension depth shrank with the wavenumber $a$ increased. At $a=14.0$, the bioconvection solution has two convection cells. Therefore, the bioconvection solution is shown with two convection cells in Fig. \ref{fig:phaseportrait.pdf}(b), which corresponds to a thermal Rayleigh number of $100$. In this scenario, an increase in the thermal Rayleigh number causes a reduction in the depth of the small convection cell that forms close to the top of the suspension ( see Fig. \ref{fig:phaseportraitRt.pdf}). A large $R_T$ means the temperature gradient is higher, which will result in the Rayleigh-B$\acute{e}$nard convection, i.e., the convection will be driven by the generated buoyancy force resultant from the temperature gradient. In other words, when $R_T$ is great enough, the Rayleigh-B$\acute{e}$nard convection will be the dominant mode of convection, which is opposed to the bioconvection, then the single convection cell can be found. As a result, a double convection cell will change into a single convection cell as the thermal Rayleigh number $R_T$ value reaches $1000$. 

The correlation between $R_a^c$ and $R_T$ is shown in Fig. (\ref{fig:15v0.5klewis.pdf}) for a variety of various values of $Le$. Here, the fixed parameters are $\hbar=0.5$, $U_s=15$, $\chi=-0.02$, and $\mathcal{G}_c=0.64$. It has been shown that the $R_a^c$ reduces as the $R_T$ increases. Here, the critical thermal Rayleigh number is $1100.5$ which is the same for all values of Lewis number $Le$. Thus, the critical thermal Rayleigh number $R_T^c$ is independent of the Lewis number $Le$, whereas the critical bioconvection Rayleigh number $R_a^c$ is strongly influenced by the Lewis number $Le$. According to the definition of Lewis number $Le$, it is a direct connection to diffusivity. On the contrary, if $R_a=0$, the solution to the system is independent of the diffusivity $D$. It would be reasonable to conclude that a significant part of the process of bioconvection is influenced by the value of diffusivity.

The impact of thermal Rayleigh number $R_T$ is described in Fig. (\ref{fig:15v1k.pdf}) for fixed parameters $\hbar=1.0$, $U_s=15$, $\chi=-0.485$, $\mathcal{G}_c=0.51$, $Le=4$. If $R_T$ is equal to zero, a single oscillating branch splits off from the associated stationary branch of the neutral curve approximately at $a=3.9$, and the oscillatory branch possesses the most unstable bioconvective solution. The critical wavenumber and Rayleigh number are $2.0$ and $61.72$, respectively. When $R_T=500$ is reached, the most stable solution is still located on the overstable branch. Overstability occurs at $a_c=2.0$ and $R_a^c=43.27$. $\gamma=0\pm 9.36$ is the pair of complex conjugate eigenvalues that corresponds to $R_T=500$. The change that can be seen in this instance is referred to as a Hopf bifurcation. The bioconvective flow patterns are mirror reflections of each other, corresponding to the complex conjugate pair of eigenvalues. One oscillation takes $2\pi/{Im(\gamma)}=0.66$ units of time to complete. The bioconvective fluid movements become nonlinear on a timeline that is much less than the duration of time that was projected to be spent in an unstable state. Therefore, it is possible to see the convection cells and flow patterns that occur throughout a single cycle of oscillation by using the perturbed eigenmodes $w^*$ and $T'$, respectively. Fig. (\ref{fig:periodic.pdf}-\ref{fig:periodicRt.pdf}) illustrate the flow pattern that is generated by the perturbed velocity $w^*$ and the perturbed temperature $T'$ during one cycle of oscillation. Here the fixed parameters are $\hbar=1.0$, $U_s=15$, $\chi=-0.485$, $\mathcal{G}_c=0.51$, $Le=4$, $R_T=500$, $R_a^c=43.38$, $a_c=2$, $Im(\gamma)=9.36$, (a) $t=15.0$, (b) $t=15.22$, (c) $t=15.44$, and (d) $t=15.66$. $\Lambda^c=3.14$ is the wavelength of the most unstable solution. It shows that the solution for the traveling wave is shifting to the left. The most unstable solution for perturbed velocity is mode one whereas mode two is for perturbed temperature. Fig. $\ref{fig:timeinv+limitcycleRt.pdf}$ depicts the time-evolving perturbed temperature component $T'$ together with its accompanying phase diagram for the value of $a^c$ equal to $2.0$. Again, it is important to point out that the period of oscillation $2\pi/{Im(\gamma)}$, is the control or bifurcation parameter, and as a result, the bioconvective flow destabilization results in the formation of a limit cycle, also known as an isolated cycle. Through the process of bifurcation analysis, the emergence of a limit cycle as a consequence of flow destabilization is once again identified as the Hopf bifurcation. This Hopf bifurcation's supercritical nature leads to a stable limit cycle, according to linear stability theory.
As $R_T$ becomes closer to $800$, a single oscillating branch separates from the accompanying stationary branch of the neutral curve approximately at the point where $a=2.9$, however, the solution that is the most unstable is found on the stationary branch. As $R_T$ approaches $931.5$, the critical Rayleigh number $R_a^c$ decreases to zero. Therefore, the suspension becomes unstable on its own, and the bioconvection can be concealed by the traditional Rayleigh-B$\acute{e}$nared convection, even though it could be bearable for certain types of microorganisms. In Fig. (\ref{fig:frequency.pdf}), the positive frequency curves for different thermal Rayleigh numbers are described. Here, the fixed parameters are $\hbar=1.0$, $U_s=15$, $\chi=-0.485$, $\mathcal{G}_c=0.51$, and $Le=4$. This illustrates exclusively the positive frequency of complex conjugate pairs of eigenvalues. Near the point where the modes overlap, the frequency approaches $0$ and then drops off completely.

\section{Conclusion}
\label{sec6}
This study is focused on the linear stability analysis of a suspension that contains phototactic microorganisms and is heated from below. The lower boundary is considered rigid, while the upper boundary is stress-free. Both stationary and oscillatory instability are encompassed for revealing intricate insights into the underlying mechanisms of bioconvection. The vicinity around the position of the maximum concentration sublayer aligns with the location exhibiting the critical total intensity. The lower region of the sublayer manifests a positive phototactic response, while the upper region displays a negative phototactic inclination. As the critical total intensity experiences an escalation, the sublayer's position shifts towards the upper surface.
Consequently, the domain of positive phototaxis expands, causing a shift in the position of maximum concentration towards the upper surface. This transition is accompanied by amplification in the value of maximum concentration. As the positive phototactic region reinforces the convective processes, the corresponding critical Rayleigh number decreases, rendering the system more susceptible to instability. The instability of the system is described by two Rayleigh numbers: first, the bioconvection Rayleigh number $R_a$; and second, the thermal Rayleigh number $R_T$, which indicates the effect of the vertical temperature gradient. It has been discovered that the critical total intensity and Lewis number have a significant impact on the critical bioconvection Rayleigh number, whereas it does not influence the critical thermal Rayleigh number. A system becomes more unstable as the Lewis number grows due to the critical Rayleigh number drops. The growth rate increases as the thermal Rayleigh number increases so that the critical bioconvection Rayleigh number decreases and the system becomes more unstable. Bioconvection will be obscured by classical Rayleigh-B$\acute{e}$nard convection if the value of thermal Rayleigh number $R_T$ is increased to the point where bioconvection Rayleigh number $R_a^c$ approaches zero. This will cause the suspension itself to become unstable, albeit being tolerated by some microbes. As the thermal Rayleigh number grows, a transition from an overstable to a stationary branch can be seen. For higher $R_T$, the Rayleigh-B$\acute{e}$nard convection will be the dominant mode of convection, which is opposed to the bioconvection, then the single convection cell can be found.

\section{Data availability}
The data that support the findings of this study are in the article. 
\section{Declaration of Interests}
The authors state that they have no conflicting interests. 
\section{Acknowledgements}
The University Grants Commission financially supports this study, Grants number 191620003662, New Delhi (India).

\nocite{*}
\bibliography{aipsamp}


\end{document}